\pdfoutput=1
\documentclass{tlp}

\submitted{5 October 2009}
\revised{2 March 2010}
\accepted{22nd November 2010}

\usepackage[pdftex]{graphicx}
\DeclareGraphicsExtensions{.pdf,.jpg,.png}
\graphicspath{{figs/}{./}}

\begin{document}
\bibliographystyle{acmtrans}

\def\eclipse{ECL$^i$PS$^e$}
\def\tkeclipse{TkECL$^i$PS$^e$}

\title{{\eclipse} - From LP to CLP}

\author[J. Schimpf and K. Shen]{\\
JOACHIM SCHIMPF\\
Monash University, Melbourne, Australia\\
\email{joachim.schimpf@infotech.monash.edu.au}\\
\\
KISH SHEN\\
Independent Consultant, London, UK\\
\email{kisshen@cisco.com}\thanks{Both authors are in part supported by Cisco Systems Inc.}
}

\pagerange{\pageref{firstpage}--\pageref{lastpage}}
\volume{\textbf{?} (?):}
\setcounter{page}{1}

\maketitle

\label{firstpage}

\begin{abstract}
{\eclipse} is a Prolog-based programming system, aimed at the
development and deployment of constraint programming applications.
It is also used for teaching
most aspects of combinatorial problem solving, e.g. problem modelling,
constraint programming, mathematical programming, and search
techniques.  It uses an extended Prolog as its 
high-level modelling and control language,
complemented by several constraint solver libraries,
interfaces to third-party solvers,
an integrated development environment and interfaces for
embedding into host environments. 
This paper discusses language extensions, implementation aspects,
components and tools that we consider relevant on the way from
Logic Programming to Constraint Logic Programming.

\noindent {\em To appear in Theory and Practice of Logic Programming (TPLP)}.
\end{abstract}


\section{Introduction}
{\eclipse} is an open source, Prolog-based programming system,
aimed at the
development and deployment of constraint programming applications.
It is also used for teaching
most aspects of combinatorial problem solving, e.g., problem modelling,
constraint programming, mathematical programming, and search
techniques \cite{AptWallace,MariottStuckey}.
It uses an extended Prolog as its 
high-level modelling and control language,
complemented by several constraint solver libraries,
interfaces to third-party solvers,
an integrated development environment and interfaces for
embedding into host environments. 

Today's {\eclipse} system has its roots in a number of other
more specialised Prolog variants that were developed in the
1980s at the European Computer-Industry Research Centre (ECRC,
a collaboration of European computer manufacturers Siemens, Bull and ICL).
These predecessor systems were
\begin{itemize}
\item ECRC-Prolog, a system that focused on efficient implementation
of data-driven execution mechanisms;
\item Sepia, a followup system with an emphasis on flexibility,
extensibility and scalability \cite{meier89};
\item CHIP, the first CLP system with a finite-domain solver \cite{dincbas88a};
\item Megalog, which emphasised persistence and database functionality \cite{BOC90};
\item Elipsys, an Or-parallel implementation of Prolog \cite{elipsys}.
\end{itemize}
{\eclipse} started in 1990 as an integration of the Sepia engine with
the Megalog database components.  In the following years, it provided
the software platform for substantial projects in the areas of
constraints and parallelism.  The result was an Or-Parallel
Constraint Logic Programming (CLP) system
with a number of constraint solving libraries, among them a
set domain solver \cite{gervet}, and
the first implementations of Constraint Handling Rules \cite{Fruehwirth}
and Generalised Propagation \cite{LeProvost92}.

In 1995, the main development activity moved to IC-Parc at Imperial College London,
where the database and parallelism work was discontinued in favour of
a stronger focus on the hybridisation of different constraint solving
techniques \cite{eclipseicl}, and this has remained a major theme until today.
Most of the Prolog extensions discussed in this paper were developed in
this period.

Subsequently, the system was exploited by Parc Technologies Ltd in the
implementation of industrial-scale applications for the airlines and telecoms
sector.  This work had implications in terms of software engineering
and programming-in-the-large, prompting the introduction of new features and
the reengineering of existing components, which we will discuss in later sections.
In 2003, {\eclipse}'s ownership transferred to Cisco Systems, and the system was
finally open-sourced in 2006, while continuing to enjoy Cisco's support.

Compared to other Prolog-based systems, we have been
relatively adventurous in {\eclipse} with the introduction of new,
mostly unpublished, language features that addressed real needs --- even
if that meant largely ignoring Prolog standardisation, which has
remained more conservative.  On the other hand we have tried not to
depart as much from the spirit of Prolog as more radical approaches like
Mercury \cite{mercury95} have done --- the strict typing and moding
approach would not fit well with the dynamicity of constraint
programming.

The organisation of this paper is as follows. Section~\ref{secprolog}
discusses how {\eclipse} implements traditional Prolog functionality
(including the module system, which plays a central role).
Section~\ref{secmodel} looks at language extensions that were introduced
largely for constraint modelling, but turn out to make Prolog
a more usable language for general programming. 
Section~\ref{seckernel} looks at kernel support for solver implementation.
Section~\ref{secmulti} gives an idea of the variety of solvers
and search components and their interaction.  The challenges of
developing large CLP based applications are addressed in section~\ref{seclarge}.

\section{Basic Prolog Implementation}
\label{secprolog}

\begin{figure}
\includegraphics[width=12cm]{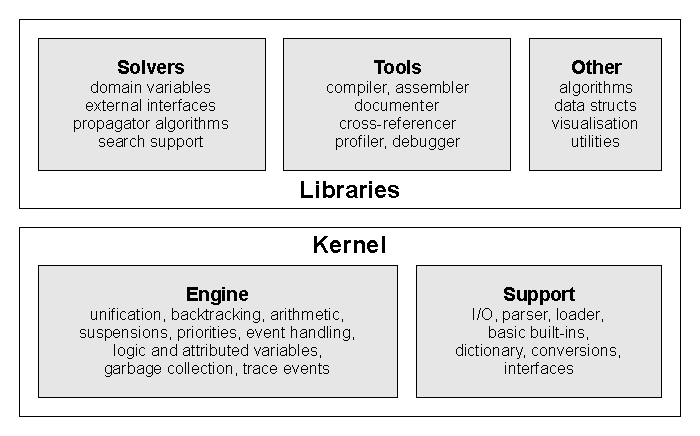}
\caption{System Structure}
\label{figlayers}
\end{figure}

Figure \ref{figlayers} gives a rough picture of {\eclipse}'s
architecture.
In this section, we briefly summarise the implementation
as far as plain Prolog functionality is concerned. We also
discuss the module system, because it provides the tools needed
to structure the rest of the system.

\paragraph{Abstract Machine:}
{\eclipse} is implemented via an abstract machine: the compiler generates
abstract machine instructions, which are then executed by a virtual machine.
The abstract machine is a variant of the Warren Abstract Machine
(WAM, \cite{warren83}), with the following main characteristics:
\begin{itemize}
\item The engine manipulates pairs of machine words (two 32-bit words,
    or two 64-bit words), called the value and the tag word.
    The main purpose of the extra word is to hold type information,
    but it is used in a few other circumstances as well (garbage collection,
    variable names, module system authentication, conversion routines).
    As opposed to single-word implementations, no tag bits are stolen
    from the value word, meaning that full pointers can be handled, and
    integers and floats (doubles in the 64-bit case) can be stored with
    their full machine precision without having to resort to a boxed
    representation on the global stack.  The obvious drawback is usually
    higher memory consumption.
\item Four separate stacks are used, called Global, Trail, Local and Control.
    As opposed to the original WAM, Local stack (containing
    environments) and Control stack (containing choice points) are split.
    This allows immediate choice point space reclamation after a cut
    or trust instruction, but has no major impact otherwise.
\item Dedicated instructions allow the creation of choice points within
    a clause. These are used for the inline-compilation of disjunctions.
\item Unification of compound terms is compiled into two isomorphic
    instruction streams, corresponding to read mode and write mode \cite{compnd}.
\item Environment slot usage is tracked via compiler-generated activity
    bitmaps. This removes the need for environment slot initialisation,
    which would otherwise be necessary for precise garbage collection.
\end{itemize}
Data is always tagged, and the following types/tags are distinguished:
four numeric types (integer, rational, float and bounded-real,
    see section~\ref{secnum}),
    with integers having 2 tags/representations (short integer and bignum);
atoms (with nil having its own tag);
strings (an atomic data type in {\eclipse});
structures (with lists having their own tag);
suspensions (section~\ref{secsusp});
handles (section~\ref{sechandle});
plain variables and
attributed variables (section~\ref{secattr}).
Further tags are used internally to label various data structures, in
particular those that are stored on the global stack, where they are
encountered by the garbage collector.

\paragraph{Compiler:}
The {\eclipse} compiler was original written in C because
compilation speed was considered of major importance.  However,
for release 6.0, a complete rewrite in the {\eclipse} language itself
was undertaken.  The main motivation for this higher level approach
was that the old compiler
had become increasingly difficult to maintain, extend and modify,
and that we wanted to incorporate some ideas from Mercury
\cite{mercury95}.
The new compiler is a modular design consisting of
\begin{enumerate}
\item The parser (the built-in predicates of the read-family).
\item The source processor (a library used by all tools that process
    source texts).
\item The actual compiler, translating one predicate at a time
    (given as a list of clauses) into symbolic abstract machine code.
\item The assembler, turning symbolic abstract machine code into a
    (relocatable) numeric representation ({\eclipse} object code).
\item The loader, which loads {\eclipse} object code into memory.
\end{enumerate}
Only parser and loader are part of the runtime system, whereas
source processor, compiler and assembler are separate libraries.
All components communicate via Prolog data structures.
Characteristics of the compiler implementation are:
\begin{itemize}
\item The compiler is implemented in {\eclipse} itself.
\item Input is a term representation of the source, or optionally a
	representation annotated with source position information,
	used for generating debugging information in the generated code.
\item Each predicate gets normalised into a single-clause form,
	i.e., the clause structure is converted into disjunctions, and
	head unifications are made explicit.
\item The compiler directly handles clauses with possibly nested disjunctions
	(forming a directed acyclic control flow graph, similar to 
        \cite{henderson1996determinism}). The retry
	and trust instructions have variants that are used when the
	clause already has an environment.  This property makes predicate
	unfolding more effective, by reducing environment allocations and
	parameter passing.
\item Inline disjunctions are indexed. Indexable variables are chosen
	by analysing the built-in predicates at the beginning of each branch.
	This is more general than just indexing on head arguments, and
	guarantees that there is no loss of indexing when a multi-clause
	predicate is unfolded into an inline disjunction.  It also provides
        a good basis for more elaborate source transformations like
        unification factoring \cite{unifact}.
\item Indexes are generated individually for every argument/variable for
	which they might be useful in some possible instantiation pattern,
	and ordered by selectivity. Selectivity is measured as the ratio
        between the number of distinct argument values and
        the number of matching alternatives.
        During execution, only one index (the most selective one for the
	actual instantiation pattern) is used.
	In our experience, this is hardly ever worse, and often
	much better than simple first-argument indexing, and it does away
	with the unnatural special status of the first argument.
	As opposed to full multi-argument indexing, this technique does
	not lead to code explosion, nor does it require extensive analysis.
        Mode declarations are taken into account to suppress unnecessary indexes.
\item  Abstract code postprocessing removes non-reachable code, reduces
       branching by duplicating short code sequences,
       eliminates indirect jumps, and generates merged instructions
       (such as multi-register moves) to speed up execution in an
       emulated setting.
\end{itemize}

\paragraph{Garbage Collection:}
{\eclipse} has garbage collection for the dictionary and the global/trail stack.
The latter is the more important, in particular in the context of
constraint processing which tends to be deterministic over long phases.
It relies on a mark-and-sweep algorithm inspired by the one developed at SICS
\cite{achs}. Because of the double-word architecture of the abstract machine,
our collector can employ a faster single-pass marking algorithm,
followed by a single-pass compaction sweep.  The double-word units make
it possible to do all the relocation work on the fly, as described elsewhere
\cite{gc90}.
Nevertheless, the compaction phase still has to scan all
unused memory, therefore an auxiliary copying collector would probably
be beneficial when the proportion of garbage is high.

A characteristic of this type of collector is that it relies on the
presence of choice points for achieving good incremental behaviour.
Long running deterministic programs can, without additional measures, 
exhibit quadratic growth in collection times and thus arbitrary slowdown.
This is due to repeated scanning of the same memory area.
One way to overcome this is to manage collections intervals carefully,
ensuring a stable ratio between newly allocated memory and the size of the
area to be scanned by the collector.
An alternative method is the creation of auxiliary choice points (which can
serve as markers for memory segment boundaries), but we have abandoned
this technique because of its undesirable interference with determinacy
assumptions across sequences of abstract machine code.

Finally, it may be worth noting that all trail cleanup is done lazily
by the garbage collector, rather than eagerly at the time of choice point
removal.  Although probably not important in practice, this guarantees
that choice point removal is a constant time operation.

\subsection{Module System}
\label{secmodule}
{\eclipse}'s module system is based on Sepia's, but was revised in 2000
in the light of previous experience.  It was felt that addressing the
shortcomings of the module system was
critical for our ability to build the multi-solver system architecture
we envisaged.  The highlights of today's system are
discussed in the following, in particular where they deviate from both
the formal \cite{ISO:2000} or the de-facto Prolog module standard.

\paragraph{Stricter Visibility Control:}
Visibility control applies not only to predicates, but to all properties
that may be attached to functors, such as goal expansions,
read-macros, portray-transformations, structure declarations, and 
global storage identifiers.
Unlike in a name-based module system,
the visibility of each functor property can be controlled separately,
rather than being linked to the functor's visibility as a whole.
In addition, there are visibility-controlled properties that are not
attached to functors, among them a module's syntax options, character class
tables, initialisation and finalisation goals.

\paragraph{Module-sensitive I/O:}
Plain Prolog already provides means to modify syntax via operator
declarations.  In {\eclipse}, there are further configurable syntax options,
I/O transformations, and character class tables. Changing such settings
will result in disaster unless their scope is clear.  They are therefore
all subject to module visibility control, and can be local or exported/imported.
This is not just a feature of the compiler: it implies that all relevant
I/O predicates are sensitive to the module context in which they are invoked.
This has proven useful for writing different
modules in different language dialects,
for defining customised syntax for data formats, and even for
reading non-Prolog languages like FlatZinc (the solver input language
that goes with the MiniZinc modelling language \cite{minizinc07}).

\paragraph{Privacy:}
Many Prolog module systems do not strictly enforce module privacy, and
allow, for instance, local predicates to be invoked from outside the module
\cite{DBLP:conf/iclp/HaemmerleF06}.  Our system allows
modules to be ``locked'', thereby limiting access strictly
to their exported interface.  This would typically be done for modules
that implement critical system functionality.
Any such protection mechanism has to preserve Prolog's meta-programming
capabilities.  Our design is built around the idea of attaching hidden
authentication tokens to module arguments, and requiring these tokens
in all built-ins that operate in the space of a locked module.

\paragraph{No static textual interface/implementation separation:}
A module's interface simply consists of the union of all its export
directives.  No textual separation is required.  Instead, tools are
provided to extract the interface information from the source or from
a loaded module.  This interface specification can then be distributed
together with the compiled abstract machine code of a module,
whenever source distribution is not an option, see figure \ref{figsp}.

\begin{sloppypar}
\paragraph{Reexport:}
\begin{figure}
\includegraphics[width=10cm]{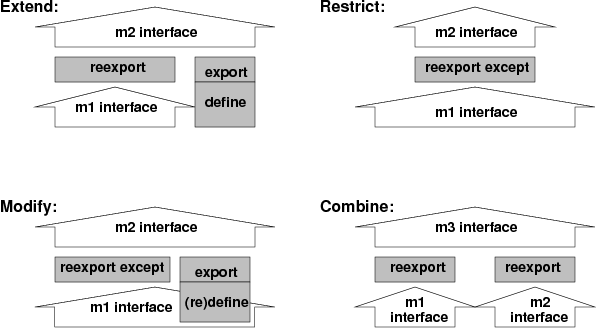}
\caption{Making modules from modules with reexport}
\label{figreexport}
\end{figure}
A basic {\em reexport} directive is defined in the ISO Prolog module standard
\cite{ISO:2000}.  We found that by introducing an additional variant
of the form \verb"reexport <module> except <items>", we could better
support the task of composing modules from existing modules, thus
giving the system some flavour of object orientation.
Figure \ref{figreexport} illustrates the main concepts:
{\em extend} the interface of an existing module by adding additional exports;
{\em restrict} the interface an existing module;
{\em modify} the interface of an existing module, by reexporting parts of
it and redefininig others;
{\em combine} functionality from existing modules by reeexporting them
from a new module.
\end{sloppypar}

\paragraph{Lookup Modules, Qualification and Name Conflicts:}
In a system with multiple constraint solving libraries, it is highly
desirable to use identical predicate names for different computational
implementations of the declaratively same constraint.  This requires
a straightforward handling of name conflicts,
which is impossible with the de-facto module standard.
Our module system implements a clear separation of the concepts of
lookup module and context module, and also allows the qualification
of a goal with multiple lookup modules.  For example
\verb"[lazy,eager]:p(X,Y)" as a shorthand for
\verb"lazy:p(X,Y),eager:p(X,Y)", invoking two different
implementations of {\em p/2}.
While in plain Prolog it would not make much sense to invoke the
(declaratively) same goal twice,
with constraint programming it can be beneficial to have several
implementations of the (declaratively) same constraint predicate
with different operational behaviours, e.g., propagators of different
strength and complexity.

\paragraph{No global items:}
No Prolog items exist globally, or outside of modules.  For instance,
built-in predicates receive no special treatment from the module system.
They are simply a set of predicates imported from a ``language'' module.
There is also no shared ``user'' module for implementation hooks.

One consequence of the above features is that it is possible to use a mixture
of different programming language dialects within a single user program.
Different user program modules can import different language modules.
Each language module will typically provide:
specific syntax in the form of operators and parsing options; 
specific semantics in
the form of predicates which may add to or replace the standard built-in
predicates; and possibly other module-local properties.
Several language modules for various Prolog dialects are provided
with the \eclipse distribution. 

Another Prolog system that has invested heavily in the module system design
is Ciao \cite{ciaomodules}.  Ciao's choices were largely motivated by the
requirements of program analysis, but it is reassuring to see that both
our groups have arrived at many of the same conclusions:
the need for local syntax, stricter and stable interface definitions,
correct semantics of module qualification, and the elimination of the
special status of built-in predicates.

\section{Language Extensions for Modelling and General Programming}
\label{secmodel}

When we started work on bringing CLP and Mathematical Programming (MP)
together, we realised that our MP collaborators did not necessarily
share our view of LP as an ideal framework for expressing constraint
models.  Being forced to express everything in terms of lists and
recursion wasn't acceptable, given that most MP models are written
in terms of arrays and quantification over index ranges.
The introduction of loop iterators and arrays was an attempt to
address these concerns, but these constructs
are useful in general programming as well.
The same is true for our structure syntax, which addresses one of
Prolog's long standing software engineering problems.
In all these extensions, we have tried to retain the spirit of Prolog
by designing them in such a way that they can be easily mapped
back into canonical Prolog.

\subsection{Arrays}
Many attempts to introduce arrays in Prolog (e.g. \cite{barklund93}),
have considered the problem of destructive updates.
This is not what we were after,
because we were more interested in declarative modelling than in
expressing imperative algorithms that rely on arrays.

Introducing pure logical arrays is not hard, and indeed,
Prolog provides them in a way. An array is an ordered collection
of items of the same type, with an index set ranging over integers
or tuples of integers, and typically constant time access to the items.
Since Prolog is dynamically typed, we can use structures as arrays,
and regard for instance \verb.wd(mo,tu,we,th,fr,sa,su). as an array constant,
or create an uninitialised array using
\verb.functor(DayArray, year, 365)..
Arguments can be accessed in constant time via {\em arg/3}, as in
\verb.arg(4, WeekdayArray, DayName)..
It is true that many early Prolog systems imposed limits on the arity
of structures, but this has become less of an issue in recent years.

\paragraph{Array Elements in Expressions:}
Using {\em arg/3} to access array elements can look clumsy, especially
when they are to be used in arithmetic expressions.  It would be so much
nicer to be able to write
\verb.Queen[I] =\= Queen[J].
instead of
\begin{verbatim}
    arg(I,Queen,Qi), arg(J,Queen,Qj), Qi =\= Qj
\end{verbatim}
This is exactly the facility we have introduced.  
Array syntax is implemented by recognising a new syntactic construct,
i.e. variable followed by list (this is backwards compatible with
standard Prolog in the sense that the new syntax does not conflict
with any previously valid syntax).
Technically, we use a trick that is familiar to Prolog implementors:
we introduce the new syntax as an alternative syntax for a particular functor.
Plain Prolog already does something similar by allowing the square-bracket
syntax for the list constructor ./2, or by allowing the '\{\}'/1 functor to
be written as a pair of surrounding braces.
We now simply define a variable immediately followed by a list as syntactic
sugar for a structure with the functor
{\em subscript/2}, with the variable becoming its first,
and the list its second argument.
For example the input \verb."M[3,4]". is parsed as \verb"subscript(M, [3,4])".
When a {\em subscript/2} term is printed, the transformation is reversed,
unless canonical representation was requested.

The second step is to allow such a term to occur as a function in an
arithmetic expression.  It is evaluated by adding a result
argument and calling the new built-in predicate {\em subscript/3}, which
is a generalised form of {\em arg/3} and
extracts the indicated array element from a possibly multidimensional array.
Like all arithmetic evaluation, this is only done in the context of
an expression,
e.g., the right hand side of {\em is/2}, or the arguments of a comparison or
other arithmetic constraint.  Normal unification is not affected, so
\verb"M=[](a,b,c), M[2]=b" will still fail, analogously to \verb"1+2=3".

\paragraph{Creating Arrays:}
To manage multidimensional arrays, represented as nested arrays,
a generalisation of {\em functor/3} is useful.
We have introduced the predicate {\em dim/2} which can be used in
two modes, either to create arrays, or to extract their dimension.
For instance:
\begin{verbatim}
    ?- dim(M, [2, 3]).
    M = []([](_341, _342, _343), [](_337, _338, _339))
\end{verbatim}
Note that we introduce here the convention of using the \texttt{[]} functor
(of arbitrary arity) for arrays.  The execution engine may in future
exploit this
by using a more efficient representation for this particular functor
(analogous to optimizations for the list functor \verb=./2=).
Observe that this choice of functor also implies that empty arrays and
empty lists look identical.

\subsection{Loops}
\label{secloops}
In the average Prolog program, the vast majority of all recursions
represent iterations.  Most of them are iterations over lists, some
are iterations over structure/array indices, and very few are
something else.
Our approach to loops has been detailed elsewhere \cite{loops02},
so we will only summarise here and point out the
usefulness for modelling, especially in connection with arrays.

The {\eclipse} loop construct {\em do/2} can be translated into an auxiliary
tail recursive predicate, plus an invocation of this auxiliary.  A call
\begin{verbatim}
    ?- ( fromto(From,In,Out,To) do Body ).
\end{verbatim}
maps into
\begin{verbatim}
    ?- do__1(From, To).
    do__1(Last, Last) :- !.
    do__1(In, Last) :- Body, do__1(Out, Last).
\end{verbatim}
Here, {\em Body} is an arbitrary, possibly complex subgoal,
{\em From} and {\em To} are terms shared with the loop's context,
while {\em In} and {\em Out} are shared with the loop {\em Body}.
The basic idea is that a simple tail-recursive predicate is generated,
where each iteration-specifier (in this example the {\em fromto}-term)
gives rise to one accumulator (one argument pair).
The intuition is that {\em First} provides the first accumulator value,
{\em Body} maps {\em In} to {\em Out}, providing an accumulator value
for the next iteration, eventually terminating when {\em Out=To}.
Importantly, arbitrarily many fromto-specifiers can be given for a single
do-loop, each of them adding one accumulator (which in the general case
requires an argument pair) to the recursive predicate.

While the above is enough to express any deterministic iterative
recursion, there are of course some very common patterns, like
iteration over list elements or integers, for which one can have
intuitive abbreviations, see Table \ref{tabiter}.

\begin{table}
\begin{tabular}{ll}
fromto(From,In,Out,To)		&general accumulator\\
foreach(Elem,List)		&list iterator and aggregator\\
foreacharg(Elem,Array)		&array iterator\\
for(I,From,To,[Step])		&integer iterator\\
param(Term)			&invariant iterator\\
\end{tabular}
\caption{Some Common Loop Iterators}
\label{tabiter}
\end{table}

The do-loop provides the functionalities of iteration, aggregation and mapping,
all of which can be combined in a single loop.
Iteration specifiers determine what is being iterated over,
termination conditions, result accumulation and fixed parameters.
In \cite{loops02}, we have argued that the proposed loop construct
provides better abstraction, better readability, shorter code and
improved maintainability compared to the equivalent recursive formulation.
At the same time, it can replace many uses of higher order operators
{\em (map, foldl)} and has advantages in those cases where it applies.
When used in the context of problem modelling, it usually has a quite
natural declarative reading in terms of quantification over lists
or arrays, or index sets.

\paragraph{Loops and Arrays:}
Loops and arrays together allow for a rather compact expression of
matrix models for constraint problems.
Figure \ref{figqueens} shows a model for the N queens problem. Note that, because
a loop introduces a local variable scope, we use the param() iterator
to indicate values that pass through the iterations unchanged.
\begin{figure}
{\small
\begin{verbatim}
    queens_array(N, Board) :-
        dim(Board, [N]),
        Board :: 1..N,
        ( for(I,1,N), param(Board,N) do
            ( for(J,I+1,N), param(Board,I) do
                Board[I] #\= Board[J],
                Board[I] #\= Board[J]+J-I,
                Board[I] #\= Board[J]+I-J
            )
        ).
\end{verbatim}
\caption{N queens constraint model with loops and arrays}
\label{figqueens}
}
\end{figure}

\subsection{Structures}
\label{secstruct}
One of the well-known concerns regarding software engineering with
Prolog is that using data structures other than lists is problematic.
The plain Prolog concept is actually rather elegant:
the functionality of structures or tuples is not provided by
a separate language construct --- instead uninterpreted function symbols assume this role.
While this simplicity is conceptually appealing, it
turns out to be a real limitation for practical programming, mainly
because structure components are identified by position only:
\begin{enumerate}
\item The programmer has to remember which positional field has which meaning.
References to numeric field positions make the code hard to maintain.
\item Whenever the structure is matched in the code, the arity of the
structure has to be known, in addition to the relevant field numbers.
\item If the definition of the structure changes, as fields are
added or removed, the programmer needs to update {\em all} occurrences
of structure templates in the source, and check all field position numbers.
\end{enumerate}
As a consequence, structures are underused in most Prolog programs.
The folklore workaround for problem 1 has been to write an access
predicate for every structure type, e.g.,
\begin{verbatim}
    employee_arg(emp(N,_,_),name,N).
    employee_arg(emp(_,A,_),age,A).
\end{verbatim}
and manipulate the structure exclusively via these access predicates,
replacing the generic {\em arg/3}.  So code like \verb"p(emp(N,A,_)) :- ..."
would have to be written as
\begin{verbatim}
    p(Emp) :- employee_arg(Emp,name,N), employee_arg(Emp,age,A), ...
\end{verbatim}
The consistent use of access predicates in lieu of pattern
matching is tedious and requires great discipline.  It also obscures the
code for the compiler: without inter-procedural
analysis, a compiler will be unable to do indexing on the argument,
since the structure no longer occurs in the clause code.
Very likely, the programmer will have to add extra cuts.
Moreover, the argument position number might be required in contexts other
than just the {\em arg/3} predicate.  E.g., in a system that provides a sorting
predicate that can sort on a structure argument, one would write
\verb"sort(2, =<, Emps, EmpsByAge)"
to sort a list of employee-structures by age.  Having such magic
numbers in the code is clearly bad practice.

\begin{table}
\begin{verbatim}
:- local struct(emp(name,age,salary)).     % Translation:
p(emp{age:A,salary:S}) :- ...          =>  p(emp(_,A,S)) :- ...
Emp = emp{salary:Sal}                  =>  Emp = emp(_,_,Sal)
arg(name of emp, Emp, Name)            =>  arg(1, Emp, Name)
sort(age of emp, =<, Emps, EmpsByAge)  =>  sort(2, =<, Emps, EmpsByAge)
update_struct(emp, [salary:NewSal],    =>  Old = emp(A1,A2,_),
                           Old, New)         New = emp(A1,A2,NewSal)
\end{verbatim}
\caption{Examples of structure syntax and their translation to canonical code}
\label{tabstruct}
\end{table}

Our solution is simply to provide syntactic sugar in such a way
that all the required patterns can be written independently of both
the structure's arity and the order and numbering of the fields.
Table \ref{tabstruct} shows some examples of this syntax.
The obvious first step is to introduce field names, which is done
via a declaration like
\verb":- local struct(emp(name,age,salary))"\footnote{
    Here, ``local'' refers to module system visibility ---
    structure declarations can be local or exported}.
This would declare a structure with name ``emp'' and
three fields called ``name'', ``age'' and ``salary''.
Then we need a better syntax for the situations where
the structure as a whole occurs in the code (be it for the purpose
of matching against an existing structure or for constructing a
new structure).
We introduce new syntax, such as \verb"emp{age:A,salary:S}",
and replace it during parsing
by the corresponding structure according to the declaration\footnote{In reality,
this is a 2-step process: the parser reads \texttt{emp\{age:A\}}
as \texttt{with(emp,[age:A])}, and a subsequent functor transformation
attached to {\em with/2} looks up the structure declaration and constructs
\texttt{emp(\_,A,\_)}.}.
The relevant structure fields are referenced by name.  Argument
positions that are not mentioned give rise to anonymous variables. 
Importantly, the \verb"{}"-syntax does not refer to the structure's arity.

For those circumstances where an argument position number is needed,
we reserve the infix operator
{\em of/2} and replace terms of the form {\em fieldname of structname}
by the field number taken from the corresponding struct declaration.
The sorting example then becomes
\verb"sort(age of emp, =<, Emps, EmpsByAge)".

As both types of replacement are done at parse time, they apply in whatever
context the constructs appear in the program.
Note that we do {\em not} propose the use of field names at runtime:
they are preprocessed away at parse time and nothing is lost in terms
of efficiency.

One remaining operation is the change of one or more structure fields,
which (in a language without destructive update) amounts to making
a new structure in which certain fields are modified while all others remain
identical.
This would normally require knowledge about all fields and their positions.
We introduce a predicate {\em update\_struct/4} that encapsulates this
knowledge: the last example in Table \ref{tabstruct} shows how an
instance of this predicate is expanded into the conjunction of two
unifications.  Again, this is usually a compile-time transformation.

Functional languages usually have syntax like {\em structure.field}
for accessing a structure field in the context of an expression.
In Prolog this is of limited use, because expressions are only
evaluated in the context of arithmetic predicates like is/2.
We have therefore not introduced a specific notation.
However, in an untyped language there is no essential
difference between a structure and an array.  We can therefore
employ our array index syntax, use the field index in its symbolic form,
and write, for instance, \verb"YearSalary is 12*Emp[salary of emp]".

To summarise, the point of our transformations is that the source code
no longer contains any mention of either the structure arity or the
position numbers of the fields.  It is therefore now possible
to simply modify the struct-declaration (reordering or adding fields)
and recompile, without having to change the rest of the program code.
The code also becomes more readable (albeit very slightly longer).


\subsection{Numbers}
\label{secnum}
In addition to standard Prolog's integer and floating point numbers,
{\eclipse} supports two further data types: {\em rationals}
and {\em bounded reals}.  They are
fully integrated into the language, can be mixed with other numeric
types in arithmetic expressions, and have their own syntax with corresponding
support in parser and term writer.  Both types can be viewed as alternatives
to floating point numbers.

\begin{sloppypar}
\paragraph{Rationals:} Rational numbers can be represented accurately
and were used in two early {\eclipse} implementations of Gauss/Simplex
solvers (by P.~Lim and C.~Holzbaur \cite{Holzbaur} respectively).
A rational is represented as normalised numerator/denominator pairs of bignums,
and written like \texttt{1\_3}.
The implementation relies on the GMP library \cite{gmp},
which is also used to provide unlimited precision integer arithmetic.
\end{sloppypar}

\paragraph{Bounded reals:} A {\em bounded real} is a safe approximation
of a real number
in the form of the closed interval between a pair of floating point bounds,
written like \texttt{0.99\_\_1.01}.
Operations on this type use safe interval arithmetic, giving accurate bounds
on the results.

The introduction of this number type was a by-product of our work on
interval constraint solvers, see section~\ref{secic}.  Its purpose may
become clearer by highlighting the difference between
a bounded real number and a variable with an interval domain.  Assume a
query succeeds in the following way:
\begin{verbatim}
    ?- p(X, Y).
    X = _{1.0..2.5}     % an interval domain variable
    Y = 1.9__2.1        % a bounded-real constant
    yes.
\end{verbatim}
This means that variable X remains unconstrained in the interval $[1.0,2.5]$,
making {\em every} value in this interval a solution.  But there is exactly
one solution for Y, guaranteed to lie in the interval
$[1.9,2.1]$, but not known more precisely.  The difference is important
in determining whether a computation is finished.

As in Prolog, numbers of the same value but different type (3, 3.0, 3\_1
and 3.0\_\_3.0) do not unify in our system.  This lack of a canonical
representation for integers has caused problems in the interaction with
constraint solvers that regard integrality as just another constraint:
the order in which constraints are propagated can
result in a variable being instantiated to an integer, or to an integral
real, and possibly lead to unexpected failures.
In hindsight, at least for the purposes of a modelling language,
having disjoint number types is probably a mistake, especially since the
usual accuracy-based arguments against merging floats and integers apply 
neither to rationals nor to bounded reals.

\section{Kernel Support for Constraints}
\label{seckernel}

One aim of {\eclipse} development was to provide an infrastructure
for research into constraint solvers.  We did not want to build
particular domains or solvers into the system kernel, but rather
develop them in {\eclipse} and deploy them as libraries.
To be able to do so, we
needed to identify concepts that are common to classes of solvers,
and implement kernel services to provide the necessary infrastructure.
The most important of these services are:
\begin{itemize}
\item flexible execution control mechanism (delayed goals, suspensions);
\item turning logical variables into constrained variables (attributed
    variables);
\item meta-programming language constructs to support these features
    (suspension handling, matching clauses);
\item module and library facilities to support clean packaging
    and the coexistence of multiple solvers;
\item robust support for compile-time preprocessing (macros, inlining, modules);
\item abstract interfaces to enable solver-independent components
    (attribute handlers, generic suspensions, constrained-condition);
\item arithmetic support (numeric types, including intervals);
\item support for interfacing external solver software (external handles
    and related trailing functionality).
\end{itemize}

\subsection{Data-driven Execution Control}
\paragraph{Coroutining:} 
One of the early attempts at improving the power of
logic programming implementations was the introduction of {\em coroutining}:
the ability to delay execution of program parts until variables
are sufficiently instantiated.  With this facility, it is possible to
turn inefficient generate-and-test programs into reasonably efficient
backtracking search programs, where tests are executed as soon as they
can be decided.  Such facilities date back at least to Prolog-II
\cite{colmerauer82} and MU-Prolog \cite{naish:thesis:86},
and were present in {\eclipse}'s predecessor systems in the
form of {\em wait} declarations (ECRC-Prolog) and {\em delay} clauses (Sepia).

Coroutining can be considered the first step towards constraint handling,
by virtue of allowing:
\begin{itemize}
\item separation of deterministic constraint setup and nondeterministic search code;
\item eager constraint-checking behaviour by waiting for sufficient
    instantiation;
\item automatic interleaving of the search process with constraint processing;
\item simple forms of propagation, such as delaying until only one variable is
    left in a goal, and then computing the variable's value\footnote{
We note that, to implement the latter technique correctly, it is
not enough to trigger execution by instantiation: it must be
possible to trigger on variable-to-variable aliasing, since this event
can reduce the number of variables in a goal.}.
\end{itemize}

\paragraph{Suspensions:} 
\label{secsusp}
To support constraint propagation more generically, we decided to reuse
the delay/wake machinery for coroutining that we had inherited from Sepia,
but to allow
additional trigger conditions for waking.  Since such conditions
are solver-specific, and solvers were meant to be definable in
libraries, we decided to separate the shared concept of a ``delayed goal'' from
the different waking conditions.
The abstract machine data type we introduced to
represent a delayed goal without waking conditions is called a {\em suspension}\footnote{To
our knowledge the name is used in SICStus with a related but different
meaning.}.

\begin{figure}
\includegraphics{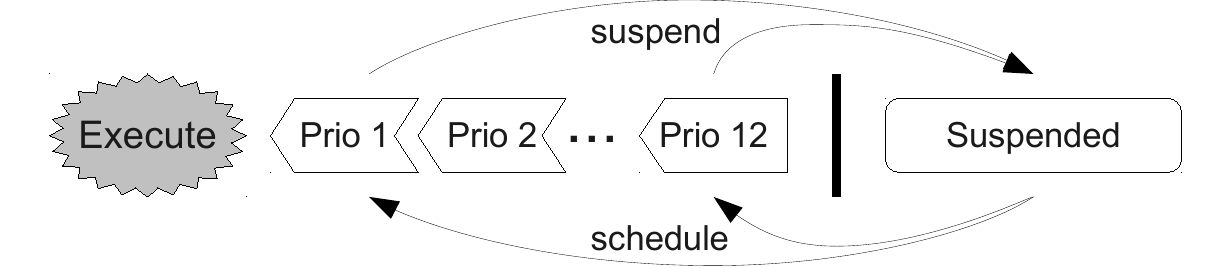}
\caption{Structure of the resolvent}
\label{figresolv}
\end{figure}
Figure \ref{figresolv} shows the structure of the resolvent, i.e.,
the collection of goals still to be satisfied.
It consists of an active part (ordered by priorities, see below)
and the currently inactive, suspended part.
Goals in the suspended part of the resolvent are represented by suspensions.
A goal enters the suspended part of the resolvent when it is
created via the {\em make\_suspension/3} built-in, analogously to the way a goal
becomes a part of the active resolvent when created via {\em call/1}.

We draw attention to the fact that our system maintains an {\em explicit}
representation of the suspended resolvent.
If a suspended goal were just a data structure stored within an
attribute, then it would be entirely the programmer's responsibility
to enforce the goal's semantics, i.e., to invoke it eventually.
If the goal were never invoked, it would be incorrectly considered true.
In our scheme, the abstract machine keeps track of each delayed
goal right from the moment it is created, independently of its attachment
to variables or trigger conditions.  Cases of unsolved subgoals
(``floundering'') are therefore always detectable.  Further advantages are
that suspensions can be manipulated via generic kernel primitives, and
that they can be displayed in a solver-independent fashion by the toplevel
and the debugger's delayed goal viewer.

\paragraph{Priorities:}
\label{priorities}
In a constraint solving system, a single event (such as updating a domain bound)
will typically wake many constraint agents (represented by suspensions)
at once.  It is helpful to have some control over the order in which they
are actually executed, since they may exhibit vastly different performance
characteristics: constraints with few variables will generally propagate faster,
linear-time propagators faster than quadratic ones, etc.  We therefore
associate suspensions with priorities, which determine the execution order
after waking.  A simple system with 12 priority levels is used.
Goals that wake up with higher priority can interrupt currently running
goals with lower priority.  High-priority goals can also be used for
tracing and debugging, and for creating data-driven animated visualisations.
Although the scheme imposes some overhead, we have found the functionality
worthwhile.
Recently, other Constraint Programming systems have also implemented
priorities \cite{mgecode09}.

\paragraph{Waking Conditions:}
The usual (though not the only) way to provide for waking of
suspensions is to associate them with conditions that occur within a
specified set of variables.  Three of these conditions are pre-defined
by the {\eclipse} kernel:
\begin{description}
\item[Instantiation] is the most obvious one: with this condition,
a delayed goal gets woken when at least one of the variables in a specified set
becomes instantiated.
\item[{\em Binding}] subsumes {\em instantiation}, but also includes aliasing of
variables. It is required in a case like the sound difference predicate
\verb"X ~= Y", in other systems known as \verb"dif(X,Y)".
As written, such a goal will delay because it is not decidable.
But unifying of X with Y should wake it and lead to failure, even without
instantiation.  When suspended under the {\em binding} condition, a goal
will wake when any two variables in the specified set are unified,
i.e., whenever the number of variables in the set is reduced.
Thus, {\em dif/2} can be written as
\begin{verbatim}
dif(X,Y) :- (X==Y -> fail ; suspend(dif(X,Y), 3, [X,Y]->bound)).
\end{verbatim}
Here, the {\em suspend/3} built-in creates a suspension of priority 3 for
\verb.dif(X,Y)., and associates as waking condition any {\em binding} 
within the variable set $\{X,Y\}$.

\item[{\em Constraining}] is unique to {\eclipse} and is an
abstract condition indicating that a variable was constrained in some way.
The concrete meaning is defined by the libraries that implement the
constrained variables.  The abstract condition makes it possible
to write generic, solver- and domain-independent tools,
such as the following predicate that eagerly prints a message whenever
a variable becomes further constrained during computation:
\begin{verbatim}
report(X) :-
    write(constrained(X)), suspend(report(X), 1, X->constrained)).
\end{verbatim}
\end{description}
\begin{figure}
\includegraphics{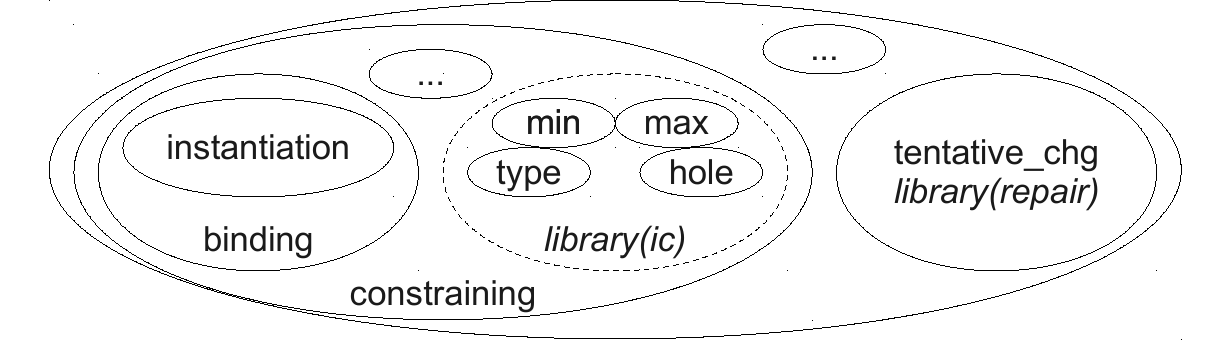}
\caption{Hierarchy of the three generic and some library-defined waking conditions}
\label{figwaking}
\end{figure}
Other waking conditions can be defined by libraries, using generic
built-ins for manipulating suspension lists and attributed variables
(section~\ref{secattr}).  Figure \ref{figwaking} shows the hierarchy of
generic conditions together with examples of library-defined ones:
the interval solver {\em library(ic)}
defines 4 waking conditions: lower and upper domain bound change,
creation of a hole in the domain, and type restriction from real to integer.
All these conditions constrain the variable further and are thus subsumed
by the {\em constrained} condition.
The {\em repair} library on the other hand implements a waking condition
called  {\em tentative\_change}, which is not considered
as constraining the variable.
Finally, it is possible to have waking conditions that are not related to
variables, but instead to certain points in execution. One such example
is the success of the subgoal in an all-solutions predicates like {\em findall/3},
where we want to make sure that no goals are left delayed.

\paragraph{Suspension States and Demons:}
\label{secdemon}
During execution, a suspension data structure may be attached to multiple
waking conditions (typically related to variables).
Since we have a two-stage waking process of (1) scheduling for priority-based
execution, and (2) actual execution from the front of the priority-queue,
we need to take care of multiple redundant waking.
This is implemented by having stateful suspensions indicating whether
they are in the suspended, scheduled, or already executed state.

Another original enhancement of the suspension system that was introduced
for the needs of constraint propagation was the concept of {\em demons}:
while a goal that simply waits for the instantiation of one variable
will only be suspended once and woken once, a goal that performs a task
like domain bound propagation may be woken many times, each time
re-suspending as exactly the same goal with the same variables.  To better
support this requirement, we introduced predicates that remain in the suspended
resolvent even after having been woken.  Declaratively, this can be viewed
as these predicates have an implicit (and thus efficient) self-recursive call.

\subsection{Implementing Constrained Variables}
\label{secattr}
{\eclipse}'s predecessor system Sepia had delay-variables, to which
delayed goals were attached by the system in an opaque way.
This was replaced by an open and more flexible mechanism, namely
attributed variables \cite{Holzbaur92}, which are
a generic way to attach (meta) information to a logical variable.
Examples of such information are:
\begin{itemize}
\item lists of goals to be woken on certain variable-related events (suspension lists);
\item unary constraints on the variable, like type or domain;
\item link to the representation of the variable in an external solver;
\item information with no effect on semantics, like debugging information or
    variable name.
\end{itemize}
We typically use coarse-grained attributes: a module (often a constraint
solver) defines no more than one attribute, but the attribute
itself will normally be a compound data structure.

Since attributes are usually meant to modify the semantics of the variables they are
attached to, they affect a range of generic operations in the basic
Prolog system, unification being only the most obvious one.
We believe that {\eclipse} is unique in the degree to which it extends 
basic Prolog semantics to attributed variables.
As soon as an attribute definition is loaded into the system, it optionally
installs handlers and hooks, which the generic system operations
can use on encountering a variable with the new attribute.
The operations whose semantics can be extended in this way are listed below.

\paragraph{Unification:}
An attribute handler is invoked immediately after an attributed variable has
been unified with a nonvariable or another attributed variable.  The handler
must first check whether the unification is allowed (by considering,
for example, the domain information within the attribute).
If so, goals associated with the variable have to be woken if their respective
waking conditions apply.  In case of variable-variable unification, a
new attribute for the resulting variable may have to be computed.

\paragraph{Unifiability and subsumption testing:}
Specialised handlers can be provided to compare domains and thus
extend the system's generic operations
for unifiability testing ({\em not\_unify/2}) and subsumption testing
({\em variant/2} and {\em instance/2}).

\paragraph{Term copying:}
This handler enables the {\em copy\_term/2} built-in to give a meaningful result
for attributed variables.  Typically, any unary
constraint (such as the domain) on the variable would be reflected in the copy.

\paragraph{Anti-unification:}
This is an interface supporting {\em Generalised Propagation} (section
\ref{secgp})
\cite{LeProvost93b} by defining its fundamental operation:
anti-unification computes the most specific generalisation of two terms,
as precisely as the expressiveness of a particular attribute allows.
For example, given the availability of finite-domain attributes,
two integers can be generalised
into a variable whose domain ranges over these two integers.

\paragraph{Constraining:}
We discussed above the generic {\em constrained} waking condition.
In order to define what it means for a particular type of attributed variable
to become ``more constrained'',
all code that implements operations on the corresponding attribute must
notify the system accordingly.
For instance, the interval constraint solver can constrain variables
by excluding domain values in various ways, and should therefore
notify the system on these occasions.

\begin{sloppypar}
\paragraph{Bounds access:}
The system defines the built-in predicates {\em set\_var\_bounds/3} and {\em get\_var\_bounds/3}
to provide a generic way to access numeric variable bounds.
The built-ins obtain their information via a handler predicate defined
together with the attribute.
This can be used for solver communication, e.g., querying the propagation
results of other solvers, or broadcasting new bounds to others.
\end{sloppypar}

\paragraph{Attribute-specific waking conditions:}
New waking conditions are made available simply by allocating a slot for
a corresponding {\em suspension list} within an attribute.
To delay a goal under the new condition, the system simply inserts
a suspension into this list.
An interval variable, for instance, has one suspension list associated
with changes to the lower bound, and one for upper bound changes.
These lists are in addition to
the pre-defined lists for instantiation, aliasing and general constraining
(figure \ref{figwaking}).
The primitive solver operations for changing variable bounds are responsible
for scheduling the goals from the appropriate list(s):  a lower bound change,
for instance,
should schedule the lower bound list as well as the constrained-list.
Once a suspension list is scheduled for execution, its member goals will
start executing according to their priorities, see section~\ref{priorities}.

\subsection{Preprocessing, or Getting term-expansion Right}
\label{sectermex}

Most Prolog systems implement the term-expansion facility.  This is a
powerful way of rewriting terms during compilation, and has many
useful applications.  However, its design is too simplistic in several
respects.  There are at least three contexts in which one may want to
transform an input term:
(1) when it occurs as a clause during compilation, (2) when it occurs as a goal
during compilation, and (3) when it occurs during general I/O
(when it is a data structure that needs to be translated
to/from some internal representation).
The traditional term-expansion mechanism makes it hard to distinguish
(1) and (2), and is not able to do (3) because it is only applied during
compilation, not term-reading in general.
Other shortcomings are the lack of cooperation with the module system
and the problem of safely combining different expansions:
term\_expansion clauses are global, and
committed to the first one that succeeds.  The clauses themselves have no
knowledge about the module context in which they occur, and thus cannot
be selective in their transformations.  Some implementations have added
goal expansions to partly address these problems.

We have opted for a different, more disciplined mechanism:
transformations are always associated with functors, and their
visibility is controlled by the module system.  Moreover, there are
three types of input transformations according to the three categories
mentioned above, plus corresponding output transformations.
The different types are described below.
Because the transformations are independent of each other,
a single functor can have more than one associated transformation.

\paragraph{Clause expansion:}
An example is the declaration for grammar rules:
\begin{quote}
\begin{verbatim}
:- export macro((-->)/2, trans_grammar/3, [clause]).
\end{verbatim}
\end{quote}
It says that whenever a clause with toplevel functor \verb'(-->)/2' is
encountered during compilation, in a module where this transformation
has been imported, it must be transformed by the transformation predicate
{\em trans\_grammar/3}.  The latter takes as arguments the original clause
plus its context module, and returns a transformed clause.  Apart from being
applied more selectively, this is similar to term-expansion.

\paragraph{Goal expansion:} Goal expansions are declared like
\begin{quote}\begin{verbatim}
:- inline(p/1, trans_p/3).
\end{verbatim}
\end{quote}
meaning that occurrences of {\em p/1} goals will be expanded using
the transformation predicate {\em trans\_p/3}.
This takes as arguments the original goal
and its context module, and returns a transformed goal.
Unlike in the other cases,
there is no visibility specification for this expansion: its
visibility is linked to the visibility of the {\em p/1} predicate, i.e., the
transformation will be applied in all modules where {\em p/1} is visible,
or even when qualified calls to {\em p/1} are made (e.g.\ \texttt{m:p(X)}).
This ensures that goal transformations always match the
corresponding predicate definitions, which is of special importance
in the case where different definitions
and different goal expansion rules for the same predicate name
co-exist in different libraries.
Goal expansions are used widely in {\eclipse},
e.g., for implementing {\em is/2}, for compile-time preprocessing of constraints,
and for the do-loop transformation (section~\ref{secloops}).

\paragraph{General input macro:} A general term macro declaration looks like
\begin{quote}\begin{verbatim}
:- local macro(foo/1, trans_foo/2, [term]).
\end{verbatim}
\end{quote}
It means that every time a {\em foo/1} term is read (even as a sub-term)
in a module context where this declaration is visible, it is transformed
by the predicate {\em trans\_foo/2}.  This transformation is done by the parser,
not only in the context of the compiler, but whenever a predicate of the
{\em read/1} family is invoked from within the right module context.
This type of transformation is
used internally to implement structure syntax (section~\ref{secstruct}).
Transformations are done in a bottom-up
fashion, so any arguments of {\em foo/1} are already transformed when {\em trans\_foo/2}
receives the term for processing.  Macros can be declared local or exported.

\paragraph{Output transformations:} The symmetric counterparts of the three
input transformations above are output transformations: they are of type
{\em clause}, {\em goal} or {\em term}, and are also associated with a functor:
\begin{quote}\begin{verbatim}
:- local portray(foo/1, trans_foo/2, <type>).
\end{verbatim}
\end{quote}
These allow an internal representation to be turned back into an
external representation before output.  Because this a term-to-term
mapping, it can be performed before arbitrary term output predicates.
This is more flexible than the traditional {\em portray/1} hook, which
produces output directly (and has rightly been omitted from the ISO standard,
but without having been replaced by a better alternative).

%


\subsection{Destructive Updates and Timestamps}
\label{secstamp}
The usefulness of attributed variables would be quite limited
without destructive updates.  Even if destructive updates were unavailable on
the language level, they would be needed for implementation-level
data structures.  Obvious examples are updating a variable's domain,
or modifying a suspension list, both stored inside an attribute.
On the abstract machine
level, these all amount to replacing one non-variable value with another
--- an operation that does not occur in pure Prolog.  To allow this in the
presence of backtracking, we have to extend the trailing mechanism, such
that it allows resetting the content of a location to an arbitrary
previous value.  This change, however, creates the new
problem of multiple redundant trailing of the same location: 
a location can be modified arbitrarily many times, but only the value
that was current when the previous choicepoint was created must
be restored.  The first published solution \cite{timestamps90} to this problem
involves keeping choicepoint-related timestamps together with the
trailed locations.  These timestamps indicate whether a location has
already been trailed since the last choicepoint was created.
In {\eclipse} we use two related techniques:
if we have control over the layout of the trailed
data structure, we add a timestamp field to it.
As the timestamp, we use the global stack pointer at the time
of the last choicepoint creation. To force this to be unique, we make
sure that at least one global stack cell is allocated along with every
choicepoint.
In case we cannot add a timestamp to the data (e.g.\ in the {\em setarg/3} predicate
which destructively updates an argument of an arbitrary Prolog structure),
we use the address of the old value as an indication of its age,
and trail only if it is older than the last choicepoint.  The new value
is forced to have an address that represents the age of the binding,
if necessary by allocating an auxiliary global stack cell and adding an
indirection.  The technique is similar to a class of techniques proposed
by Noy{\'e} \cite{noye94}.

\subsection{Interfacing External Solvers}
\label{seclowlevel}
We have successfully connected external solver libraries to {\eclipse},
such as the mathematical programming system {\em COIN-OR} \cite{coinor} and
the constraint library {\em Gecode} \cite{mgecode09}.
To be efficient enough, these interfaces must be low-level. They are typically
written in C/C++, use dynamically linked libraries, and require direct
access to solver data structures on one hand, and {\eclipse}'s abstract
machine data structures on the other.  They are supported by the following
kernel features.

\paragraph{Low Level Programming Interface:}
This interface allows direct access to the abstract machine's data
representation.  Apart from interfacing external solvers, it is
also used for connecting other software, such as databases,
or to implement procedural algorithms more
efficiently than would be possible in Prolog.

The interface exposes a subset of the operations used to implement the
{\eclipse} runtime system itself, and consists of macros, type definitions
and interface functions. It is powerful and efficient, but 
requires detailed knowledge about the internal architecture and concepts.
A low-level interface exists for the
C programming language and, with wrapper classes, for C++.
It is bi-directional in that it enables the implementation of external
predicates in C/C++, but also allows {\eclipse} goals to be
constructed and executed from C/C++.

\paragraph{External Data Handles:}
\label{sechandle}
Recurring problems in interfacing general software to a Prolog-like
system are the handling of backtracking and garbage collection.
While Prolog data structures are discarded from the stacks on backtracking,
or removed by the garbage collector when they are no longer accessible,
the same has to be arranged explicitly for data allocated by interfaced
software.  We achieve this by having a special Prolog-side data type
(called a {\em handle}) which refers to the external data.  In addition,
every handle is associated with a method table which lists methods
specific to the data that is being pointed to, among them a method for
releasing the storage.
When the handle is discarded, the external object is automatically freed. 

\begin{figure}
\includegraphics[width=8cm]{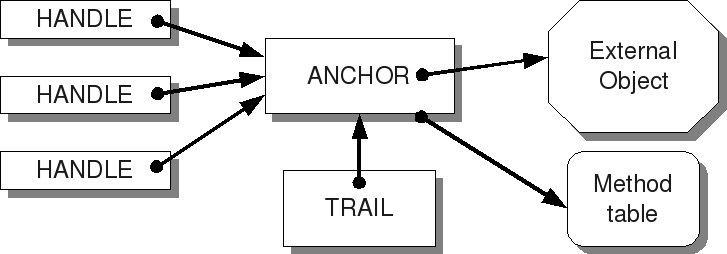}
\caption{External Data Handles}
\label{fighandle}
\end{figure}

Handles cannot be simple tagged pointers directly to external data,
because the Prolog abstract machine will blindly make copies of such
tagged pointers, making it difficult to keep track of whether the data
is still referenced.  Our solution is to introduce an indirection: the
external data is referenced only once from a dedicated global stack
cell (called an {\em anchor}), which in turn can have arbitrary references
from other Prolog objects (Figure \ref{fighandle}).  When the garbage collector detects that the
global stack anchor has become garbage, it invokes the external object's
free-method.  The object must also be freed when the anchor is
popped on backtracking --- we achieve that with a special trail entry
that, on backtracking, leads to invocation of the free-method.
Should the anchor become garbage before backtracking, the trail entry
becomes redundant and is removed by the garbage collector together with
the anchor itself.

Apart from deallocation, we also need to consider the case where the external
object gets modified in the course of the computation.  If the Prolog
side backtracks to a state before the external modification was made,
the modification will typically have to be undone.  We do this by
trailing pointers to user-defined C/C++ ``undo-functions'', which
will then be invoked on backtracking.  As with the trailing of destructive
updates, the technique has to be combined with a timestamping mechanism
to be scalable.

\section{Library Examples}
\label{secmulti}

For the constraint application programmer,
working with {\eclipse} involves:
problem {\em modelling} using an extended Prolog;
choosing {\em solver} libraries appropriate for particular problem domains;
considering {\em libraries} for generic techniques, or for specific
solver hybridisation methods;
implementing search {\em heuristics}, solver cooperation, or problem-specific 
propagation by using {\eclipse} as a programming language.
This section presents some typical libraries that the system provides as
building blocks to support these tasks.

\subsection{A ``Native'' Solver: Interval Constraints}
\label{secic}
The interval solver {\em library(ic)} provides unified handling
of continuous and integer domains. Its conceptual computation domain
is the real numbers, plus infinities.
Numbers can be constrained to be integral, and constraints
can range over a mixture of integral and non-integral variables.
A wide range of constraints is supported, including linear and
nonlinear arithmetic operations, and a number of symbolic constraints
such as {\em alldifferent/1}.
The functionality subsumes that of a finite domain solver.
The code in figure \ref{figqueens} uses this library.

The solver is implemented natively, with much of the code
written at the {\eclipse} language level.
Interval variables are implemented as attributed variables, and their
bounds are represented as a pair of floating point numbers.
The kernel's interval arithmetic is used, keeping
rounding errors under control.
Integer variables can have additional bitmaps to represent holes in their
domain, and bitmap operations are accelerated using functions interfaced through
the low-level C interface.

Most of the constraints are implemented using AC-3 style propagators
\cite{mackworth77}, which recompute domains after changes.
The propagators themselves are simply delayed goals with suitable waking conditions.
The solver defines new waking conditions appropriate for its domain
variables: {\em min} (lower bound change), {\em max} (upper bound change),
{\em hole} (non-bound domain reduction), and {\em type} (imposition of integrality).
As the general mechanism of attributes and suspensions is used for 
implementing constraint behaviour, there is no need for additional low-level
support.
The following illustrates how to implement
{\tt geq(X,Y)}, a simple $X \geq Y$ constraint, where
{\tt X} and {\tt Y} are variables or integers:
\begin{verbatim}
   geq(X, Y) :-
       ic:get_max(X, XH), ic:get_min(Y, YL),
       ic:impose_min(X, YL), ic:impose_max(Y, XH),
       ( var(X),var(Y) -> suspend(ge(X,Y), 0, [X->ic:max,Y->ic:min])
       ; true ).
\end{verbatim}
We use a suspension that wakes when either the upper bound of {\tt X} or the
lower bound of {\tt Y} is narrowed. Any change in bound is
propagated to the other variable using library primitives: 
the value for the bound
that may have changed is obtained by {\em get\_max}/{\em get\_min},
then that bound is imposed on the other variable using
{\em impose\_min}/{\em impose\_max}. The interested
reader is referred to the documentation provided with {\eclipse} for more
details.

As no hidden mechanism is used, and assuming the solver exports a small
number of fundamental primitives, such as access to domain bounds,
a user can implement additional constraints on the {\eclipse} level.
This is particularly interesting given the large number of potentially
useful ``global'' constraints \cite{catalog2005},
and we have been fortunate enough to receive external contributions of such constraints,
packaged as {\eclipse} libraries for distribution.

\begin{sloppypar}
The interval solver also makes extensive use of the preprocessing facilities
(goal expansion, section~\ref{sectermex}) for
compile-time transformations of constraints.  For example, we
normalise arithmetic expressions and expand the constraint
\verb"X#>=5*(X+Y)+2" into its internal form
\verb"ic:ic_lin_con(6, 1, [2*1, 5*Y, 4*X])".
When printed, the internal form is translated back into readable form
via an output transformation.
\end{sloppypar}

\subsection{An External Solver Interface: Eplex}
The motivation for interfacing to an external solver comes from the wish to
take advantage of existing software:
comparisons have shown that a state-of-the-art
Mathematical Programming (MP) solver can be 1-3 orders of magnitude
faster than one purpose-written for CLP systems \cite{eplex05}.
Such external solvers typically provide an API in a
popular imperative language such as C/C++.

{\eclipse}'s {\em library(eplex)} \cite{eplex05} is a common interface
to several state-of-the-art MP solvers, such as
CPLEX (www.ibm.com), Xpress-MP (www.fico.com)
and COIN-OR \cite{coinor}.
It allows the optimisation of linear constraints over continuous
and integer variables by an external solver. The simplest mode of use consists
in modelling a problem in {\eclipse}, passing it to the external solver,
and returning the results.
But more importantly, the interface allows a tight integration of the external
solver's operation with the Prolog side's data-driven propagation and backtracking-based
search framework. Each MP problem can then be regarded as
being represented by a single compound constraint,
and problem solving can be triggered in a data-driven way.
A problem can be repeatedly modified (by adding more constraints to it,
and/or updating the variable bounds) and re-solved, with backtracking returning
a problem to its previously state. 

The {\em eplex} library is written in both {\eclipse} and C, using the
low-level interface described in section~\ref{seclowlevel}. 
Attributed variables and suspensions are used to provide the
constraint-like data-driven behaviour: a demon suspension which
invokes the MP solver is created, and woken whenever
the specified triggering conditions are met. The MP solver is represented 
by an external data handle, and each
{\eclipse} variable involved in an MP problem is linked to the
solver through its attribute.
The interface is fully dynamic:
any change made to a problem after setup (e.g.\ adding constraints,
changing variable bounds), is reflected in the external solver.
To maintain the logical behaviour of the whole system, any such changes
are undone on backtracking.  Implementation-wise, this relies heavily
on our trailing and timestamping facilities
(sections \ref{sechandle} and \ref{secstamp}).

\subsection{A Higher-Level Technique: Generalised Propagation}
\label{secgp}
The Generalised Propagation solver {\em library(propia)} \cite{LeProvost93b}
interprets program annotations and extracts deterministic
information from arbitrary disjunctive sub-problems. It is very
useful for prototyping unusual and problem-specific constraints, that
would otherwise need extensive reformulation into standard constraints.
It is an example of a library that relies purely on the generic system
interface to attributed variables (the concepts of constrainedness and
generalisation), and can therefore cooperate with any domain-oriented
solver.

\subsection{An Orthogonal Paradigm: Repair-Based Search}
The {\em repair} and {\em tentative} libraries implement techniques
that differ radically from the framework of domain solvers, being rather
closer to Local Search techniques:  tentative values are attached to
variables, and the amount of constraint violation is measured.
By varying the tentative values, a local search procedure can reduce
constraint violations, and find better solutions \cite{cometbook}.
There are many ways
of combining this with constraint propagation and tree search,
one successful example being {\em unimodular probing} \cite{HaniProbe}.
Interestingly, we were able to implement this paradigm using the same
underlying techniques as the other solvers.
We use attributes to attach tentative values to variables,
and we are able to use attribute handlers and suspended demons
to update violation counts, conflict sets, and tentative invariants
in an incremental fashion.  The common architecture facilitates the
implementation of hybrid schemes that combine propagation with Local Search.

\section{Programming Larger Applications}
\label{seclarge}

{\eclipse} has been used to implement a number of large applications,
many involving constraint solving. Such applications are characterised by:
\begin{description}
\item[Size:]
    Typically moderately large amounts of {\eclipse} code, some of it
    concerned with actual problem modelling, but much of it performing
    general data processing tasks: hundreds of predicates, dozens of modules,
    tens of thousands of lines of code.  Although this does not reach the
    dimensions of very large industrial software (partly due to the greater
    compactness of Prolog code), it goes beyond what is common in academic
    use, and highlights plain Prolog's limitations with respect to larger scale
    software engineering.

\item[Interfacing requirements:]
    Interfacing with a software environment, e.g., retrieving data
    from a database, producing results in the form of web pages,
    interacting via graphical user interfaces.  Frequently, such
    requirements also come in rather arbitrary form, such as ``must
    be a Java application''.

\item[Quality requirements:]
    Code must be designed, written, tested, documented and maintained
    to certain standards.
\end{description}
These issues are in part addressed by the language extensions
we have discussed earlier, such as the module system
(section~\ref{secmodule}) and data structure declarations
(section~\ref{secstruct}).
Our approach to addressing the host software interfacing requirements
involves a high-level, language-independent communication scheme that
has been described elsewhere in detail \cite{interface02}.
The main ways to achieve code quality are through training,
methodology and tools, which we review in the following.

\paragraph{Methodology:}
Solving large-scale combinatorial optmisation problems presents
additional challenges, as compared to standard software development.
ESPRIT project 22165 (CHIC-2), in which {\eclipse} served as a platform,
produced a high-level methodology \cite{Gervet01largescale}.
Concrete technical development guidelines were formulated by Simonis
\cite{HelmutMethodology03}. These build on more basic training
and tutorial material, such as \cite{EclipseTutorial03,AptWallace,HelmutELearn}.

\paragraph{Development Environment:}
Apart from supporting the build process and interactive execution,
the development environment provides tools that give
information about the state of an executing program.
The main ones are the tracer and the data inspector.

The tracer combines the classical port-oriented box model \cite{byrd80}
(enhanced with goal stack display and filtering capabilities)
with source-oriented viewing and breakpointing facilities.
The tracer's architecture is layered: during program execution,
low-level trace events are generated by the abstract machine emulator and
combined with debug information that the compiler has inserted into the code.
A second layer maps the low-level events into box-model events and reconstructs
a full call stack.
A third layer presents this information via a user interface.

Whenever execution is halted, the current state can be inspected
through a tree-browser that allows to traverse and display
all data structures associated with the current goal or its ancestors.
This tool has proven indispensable when dealing with
complex nested data structures in large programs. 
With coroutining and constraints, an additional important tool is the
delayed goals viewer, which displays the suspensions and their state.

The debugging tools have a choice of user interfaces: a traditional
command-line interface, as well as GUIs in Tcl/Tk and in Java.
The tools are independent from the rest of the development environment,
and can be attached any running (even embedded) {\eclipse} engine via
a stream-based protocol.


\paragraph{Structured Documentation:}
We support structured {\em comment/2} directives as a way to formally add
documentation to source code.  These directives can relate to a whole
module, to predicates or to data structures.  For instance, for predicates
the comment directive contains fields like: a detailed description,
mode information, summary, arguments, example usages, etc. 
Although devised independently, our solution is similar to the LPdoc system
of Ciao Prolog
\cite{lpdoc2000}
in that the documentation is provided in the form of
directives. One difference is that we do not define our
own mark-up language for formatting text, but rely on common HTML format.

\begin{figure}
\includegraphics{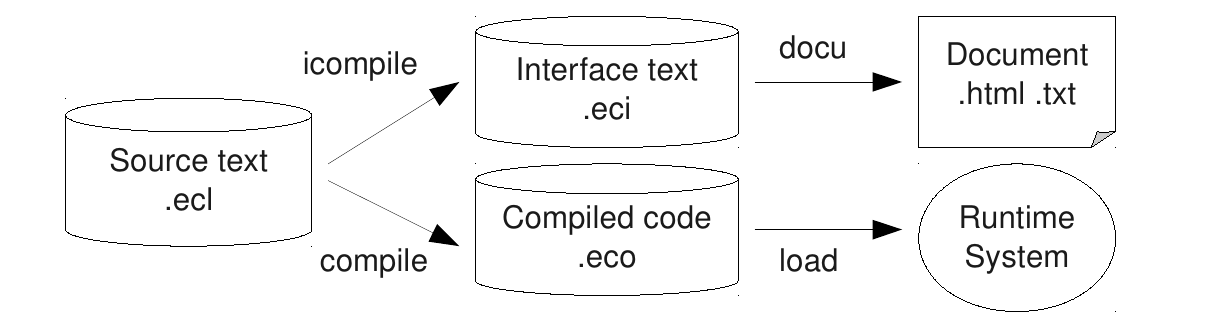}
\caption{Source Processing}
\label{figsp}
\end{figure}

Comment directives are processed in two steps: first they are 
extracted from the source file by the {\tt icompile} tool, together
with other directives that describe the module's exported interface.
The information is put into an {\eclipse} interface information (eci) file.
The rationale for this is that this file can be distributed together with
a precompiled (eco) file in place of the module source code (Figure~\ref{figsp}).
In a second step, the {\tt document} library tools process the
information in the eci file to produce reference documentation for
modules, for example in the form of HTML pages with indexes and cross-links.

\paragraph{Unit Testing and Code Coverage:}
The {\em test\_util} library provides support for unit testing.
It allows to write simple rules relating a goal with an expected outcome.
This library was initially developed to support the daily automatic test and
build of {\eclipse} itself, but has since been used to test application programs
as well.
It is supplemented by a code {\em coverage} tool that displays how
frequently each code point was executed during testing. In this way, full
test coverage can be ensured.

\paragraph{Profilers and Instrumentation:}
For performance tuning, we have developed a number of tools:
a timing profiler based on
sampling the abstract machine's program counter, which works with
fully optimised code and displays a flat
profile of the predicates in which time was spent.
Another profiler is built on top of the infrastructure for the box
model tracer; it creates a profile in terms of transitions through
box model ports, and needs the code to
be compiled in debug mode.
An even more general library provides code instrumentation by source expansion,
and can be used for analysing specific resource usage, in particular memory.

\section{Conclusion}

In the long history of {\eclipse}, many good ideas were incorporated,
but quite a few bad decisions were taken as well.  Many of them were
revised later, although this might not be surprising given the
lifespan of the system.  The usual lessons regarding software
engineering apply, in particular those about defining clean interfaces
and allowing for components to need replacement over time.

Only few of the commercial applications developed with
{\eclipse} have been documented in accessible publications.
However, open-sourcing has enabled the user community to contribute.
The contributions so far have been of high quality and, as expected,
largely in the form of libraries.  We hope very much that this trend
will continue.

There are many projects for the future which cannot be listed
here --- a large system like {\eclipse} always has construction sites.
A quite substantial but worthwhile job would be to revive the parallel
version of the system, which was mothballed almost 15 years ago.
On the language level, we want to make the system easier to use for
constraint problem solvers who don't want to know about the intricacies
of Prolog.  We also plan to continue our successful strategy of
interfacing third party solver software, and to strengthen {\eclipse}'s
role as a glue system.

We hope that our past work has been original and influential
in the wider Prolog community.  We also hope
that we have played some role in demonstrating the benefits of Logic
Programming to a wider audience in the world of optimization and
decision support.

\section*{Acknowledgements}
Owing to its long history, there are dozens of contributors to thank
for their work on {\eclipse} and its predecessor systems.
The authors would especially like to acknowledge Micha Meier, who
was the technical lead for the first 5 years (after leading the Sepia
project prior to that), and Mark Wallace who led and shaped
the project subsequently at IC-Parc, and co-wrote the book \cite{AptWallace}.
For making the whole endeavour possible in terms of
funding and environmental stability, thanks are due to Herv{\'e}
Gallaire at ECRC, Barry Richards at IC-Parc, Parc Technologies
and Crosscore, and most recently Hani El-Sakkout and Fred Serr at
Cisco Systems.

While space restrictions do not allow us to honour the individual
contributions, we extend our thanks to our many past collaborators
(alphabetic, with apologies for any omissions):
A.~Aggoun, K.~Apt,
F.~Azevedo,
J.~Bocca,
P.~Bonnet,
S.~Bressan,
P.~Brisset,
A.~Cheadle,
D.~Chan,
M.~Dahmen,
P.~Dufresne,
A.~Eremin,
E.~Falvey,
T.~Fr\"{u}hwirth,
C.~Gervet,
H.~Grant, P.~Kay,
W.~Harvey,
A.~Herold,
C.~Holzbaur,
L.~Li,
V.~Liatsos,
P.~Lim,
S.~Linton, I.~Gent,
G.~Macartney,
D.~Miller,
S.~Mudambi,
S.~Novello,
B.~Perez,
K.~Petrie,
T.~Le~Provost,
E.~van~Rossum,
A.~Sadler,
H.~El-Sakkout,
J.~Singer,
H.~Simonis,
P.~Tsahageas,
R.~Duarte~Viegas,
D.~Henry~de~Villeneuve,
N.~Zhou.
ECLiPSe also includes open source code by
R.~O'Keefe, J.~Fletcher, H.~Spencer, the GMP project, and the Mercury project.
Finally, we express our thanks to the anonymous reviewers of this paper
for their helpful suggestions.

\bibliography{sepiachip}

\end{document}